\documentclass[pra,showpacs,showkeys,preprintnumbers,amsmath,amssymb,superscriptaddress,twocolumn]{revtex4-1}
\usepackage[T1]{fontenc} 
\usepackage{makeidx} 
\usepackage{graphicx} 
\usepackage{dcolumn} 
\usepackage{array} 
\usepackage{amssymb} 
\usepackage{amsmath}
\usepackage{textcomp}
\usepackage{rotating}
\usepackage{wasysym}
\usepackage{multirow}
\usepackage{subfigure}
\usepackage{color}
\usepackage{eucal}
\usepackage{mathrsfs}
\usepackage{units}
\usepackage{rotating}
\usepackage[all]{xy}
\usepackage{float} 
\usepackage{amsmath} 
\usepackage{amsfonts} 
\usepackage{bm} 
\usepackage[amssymb]{SIunits}
\usepackage{ulem}
\usepackage{epstopdf}

\usepackage{color}

\usepackage{float} 
\usepackage{amsmath} 
\usepackage{amsfonts} 
\usepackage{bm} 

\begin{document}

\title{Influence of the external pressure on the quantum correlations of molecular magnets}

\author{C. Cruz\footnote{clebsonscruz@yahoo.com.br}} \affiliation{Instituto de F\'{i}sica, Universidade Federal Fluminense, Av. Gal. Milton Tavares de Souza s/n, 24210-346 Niter\'{o}i, Rio de Janeiro, Brazil.} 
\author{Á. S. Alves}\affiliation{Projeto F\' isica no Campus, Laborat\' orio de F\' isica de Materiais, Departamento de F\' isica. Universidade Estadual de Feira de Santana, Feira de Santana, Bahia, Brazil}
\author{R. N. dos Santos}\affiliation{Projeto F\' isica no Campus, Laborat\' orio de F\' isica de Materiais, Departamento de F\' isica. Universidade Estadual de Feira de Santana, Feira de Santana, Bahia, Brazil}
\author{D. O. Soares-Pinto} \affiliation {Instituto de F\'{i}sica de S\~{a}o Carlos, Universidade de S\~{a}o Paulo, CP 369, 13560-970, S\~{a}o Carlos, SP, Brazil}
\author{J. C. O. de Jesus}\affiliation{Projeto F\' isica no Campus, Laborat\' orio de F\' isica de Materiais, Departamento de F\' isica. Universidade Estadual de Feira de Santana, Feira de Santana, Bahia, Brazil}
\author{J. S. de Almeida}\affiliation{Instituto de F\' isica, Universidade Federal da Bahia, Salvador, Bahia, Brazil}
\author{M. S. Reis} \affiliation{Instituto de F\'{i}sica, Universidade Federal Fluminense, Av. Gal. Milton Tavares de Souza s/n, 24210-346 Niter\'{o}i, Rio de Janeiro, Brazil.}

\date{\today}

\begin{abstract}
The study of quantum correlations in solid state systems is a large avenue for research and their detection and manipulation are an actual challenge to overcome. In this context, we show by using first-principles calculations on the prototype material KNaCuSi$_{4}$O$_{10}$ that the degree of quantum correlations in this spin cluster system can be managed by external hydrostatic pressure. Our results open the doors for research in detection and manipulation of quantum correlations in  magnetic systems with promising applications in quantum information science.
\end{abstract}
\keywords{First principles, Quantum correlations, Entanglement production and manipulation}
\maketitle

\section{Introduction}

Quantum correlations play an important role in quantum information science as a remarkable resource in quantum  information processing \cite{nielsen,vedral2,horodecki,tufarelli}. The existence of pure quantum correlation has been usually inferred by the presence of entanglement. Nevertheless, although quantum entanglement provides a way to find out pure quantum correlations, it does not encompass all quantum correlations of the system \cite{cruz,yuri,yuri2,vedral,vedral4,vedral5,vedral6,modi,gu,sarandy,adesso,adesso2,girolami,girolami2,luo}. Nowadays, the notion of quantum correlations has been greatly expanded; and the measure of quantumness of the correlation has been named as \textit{quantum discord}. The study of quantum discord has been attracting a considerable attention due to their important role in many quantum information processing even when the entanglement is absent \cite{sarandy,cruz,yuri,yuri2,tufarelli}

Despite of much effort by the scientific community, the characterization of quantum correlations consist in a rather complicated task, theoretically and experimentally speaking  \cite{cruz,liu,adesso,adesso2,girolami,girolami2,luo,huang2014computing}, specially in condensed matter systems; since this difficulty increases with the number of constituents of the system. This fact has stimulated alternative measurements of quantum correlations, allowing a better control of these quantum properties in these systems  \cite{cruz,liu,adesso,adesso2,girolami,girolami2,luo,yuri,yuri2}.

In the past few years, it was understood that quantum discord can be evaluated through the measurement of some thermodynamic properties of solids, such as magnetic susceptibility \cite{cruz,yuri,yuri2}, internal energy \cite{yuri,yuri2}, specific heat \cite{cruz,yuri,yuri2} and difractive properties of neutron scattering\cite{liu,cruz2016quantum}. Recently, it has been shown that quantum discord can also exist at higher temperature, for instance, at thousands of kelvins above room temperature \cite{cruz}, showing that quantum correlations are related to significant macroscopic effects, allowing a better control of quantum correlations in solid state systems by means of materials engineering \cite{cruz}.

In this context, we show in the present work  that the degree of correlation in a spin cluster system can be affected by the structural parameters, by applying external hydrostatic pressure. We performed first-principles calculations to investigate the dependence, under external pressure, of the magnetic coupling constant of the metal-silicate framework KNaCuSi$_{4}$O$_{10}$ \cite{brandao2009magnetic} - a Heisenberg dimer on a $d^9$ electronic configuration; from which we obtain the entropic quantum discord and the entanglement of formation as a function of its magnetic susceptibility. Our results show that it is possible to manipulate the degree of quantum correlation in a magnetic system inducing a structural contraction by applying an external pressure. 
This leads to a better management of the quantum properties of these systems and opens the doors for  experimental and theoretical research of quantum correlations via first-principles, leading to a better understanding of these quantum properties with promising applications in emerging quantum technologies.

\section{Pairwise quantum correlations in a prototype material}

Our prototype material in which we investigate the influence of an external pressure on the quantum correlations via first principles calculations is the metal-silicate framework KNaCuSi$_{4}$O$_{10}$ \cite{brandao2009magnetic}. This compound is synthetic analogs to a  natural occurring minerals litidionite, an Heisenberg dimer in a $d^9$ electronic configuration and therefore an ideal realization of a two qubit system (spin $1/2$ dimer)  ruled by a Heisenberg-Dirac-Van Vleck Hamiltonian $\mathcal{H}=-J \vec{S}_1\cdot \vec{S}_2$  \cite{mario,sarandy}, where $J$ is the magnetic coupling constant \cite{mario}. The magnetic susceptibility of this system satisfies the Bleaney-Bowers equation \cite{mario,bleaney}:
\begin{equation}
\chi (T)=\frac{2 N(g\mu_B)^2}{k_B T}\frac{1}{3+e^{-{J}/{k_B T}}}
\label{eq:4}
\end{equation}
where, $g$ is the Landé factor, $\mu_B$ is the Bohr magneton, $k_B$ is the Boltzmann constant and $N$ is the number of dimers.

As calculated in reference \onlinecite{cruz}, it is possible to write the pairwise correlation function of the Heisenberg dimer as a function of its magnetic susceptibility at finite temperature
\begin{equation}
c(T)=\frac{2 k_B T }{N(g\mu_B)^2}\chi (T)-1
\label{eq:5}
\end{equation}
Hence, with $c(T)$, one can easily obtain the quantum correlations of KNaCuSi$_{4}$O$_{10}$, as a function of their magnetic susceptibility  \cite{luo,yuri,yuri2}.

The total amount of correlation in the system which is identified by the mutual information  $\mathcal{I}(\rho_A:\rho_B)=S(\rho_A)+S(\rho_B)-S(\rho_{AB})$ can be splitted into the quantum part $\mathcal{Q}$ and the classical ones $\mathcal{C}(\rho_{AB})$; where $S(\rho_{AB})=- \mbox{Tr} \left[\rho_{AB}\log_2 \rho_{AB} \right]$ is the von Neumann entropy  \cite{zurek,vedral4,liu,ma,adesso2,adesso,cruz,sarandy,paula,sarandy3,luo,datta,vedral}.
The amount of genuinely quantum correlations, called \textit{quantum discord} can be defined as the difference between the total and the classical correlation, $\mathcal{Q}(\rho_{AB})=\mathcal{I}(\rho_A:\rho_B)-\mathcal{C}(\rho_{AB})$. This difference is due to the quantum effects on the correlation between the subsystems $A$ and $B$. Thus, the entropic quantum discord depends on the magnetic susceptibility of the compound as:
\begin{widetext}
\begin{eqnarray}
\mathcal{Q}(T)=\frac{1}{4}\lbrace \left[2-3\alpha T \chi (T)\right]\log_2 \left[2-3\alpha T\chi (T)\right]+3\alpha T\chi (T)\log_2 \left[\alpha T\chi (T)\right]\rbrace - \frac{1}{2}\lbrace\left[1+\vert \alpha T\chi (T)-1\vert\right]\times\nonumber \\
\times\log_2 \left[1+\vert \alpha T\chi (T)-1\vert\right]+ \left[1-\vert \alpha T\chi (T)-1\vert\right]\log_2 \left[1-\vert \alpha T\chi (T)-1\vert\right]\rbrace& \label{eq:6}
\end{eqnarray}
\end{widetext}
where $\alpha=2 k_B / N(g\mu_B)^2$ \cite{cruz}.


Furthermore, in order to quantify the amount of entanglement in these spin system and make a comparison with the measurements of quantum discord, we adopt the measurement of entanglement of formation defined by  \cite{wootters,hill}
\begin{eqnarray}
\mathbb{E}=-\Lambda_{+}\log_2 \left(\Lambda_{+}\right)-\Lambda_{-}\log_2 \left(\Lambda_{-}\right)
\label{eq:9}
\end{eqnarray}
where
\begin{equation}
\Lambda_{\pm}=\frac{1\pm\sqrt{1-\mathbb{C}^2}}{2}
\end{equation}
and $\mathbb{C}$ is the concurrence \cite{wootters,hill,nielsen} that is written as a function of the magnetic susceptibility as \cite{cruz,yuri,yuri2}:
\begin{equation}
\mathbb{C} =\left\{
\begin{aligned}
-\frac{1}{2}\left[ 2+3\frac{2 k_B T }{N(g\mu_B)^2} \chi (T)\right]  \qquad T< T_e\\
0     \qquad T\geq T_e
\end{aligned}
\right.
\label{eq:10}
\end{equation}
where 
\begin{equation}
T_e\approx 0.91\vert J\vert /k_B
\label{temp_ent}
\end{equation}
is the temperature of entanglement, the maximum temperature below which there is entanglement in the system \cite{diogo,mario2}.

\section{First-Principles Calculations}

We performed first-principles calculations to investigate the dependence of the magnetic coupling constant of metal-silicate framework KNaCuSi$_{4}$O$_{10}$ \cite{brandao2009magnetic} under external pressure in order to evaluate this influence on the quantum correlations obtained as a function of the magnetic susceptibility of this compound.
\\
\subsection{Technical details}

\begin{figure}[htp]
\centering
\includegraphics[scale=0.43,angle=270]{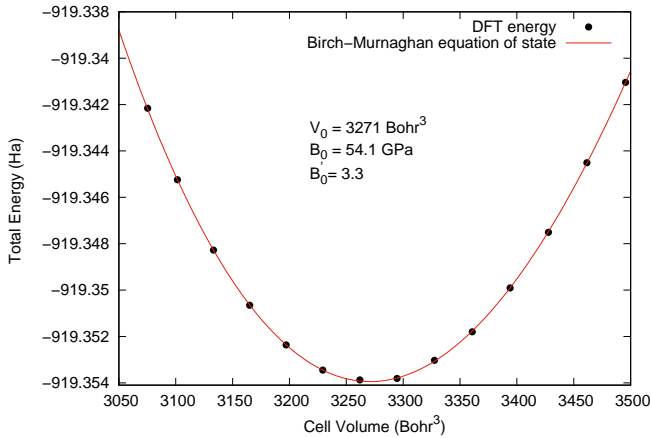}
\caption{(colour online) Total energy versus cell volume for KNaCuSi$_{4}$O$_{10}$ calculated by DFT and fitted using the third-order Birch-Murnaghan equation of state.}
\label{fig:dft}
\end{figure}

We approached this problem by using the density functional \cite{hohenberg1964inhomogeneous} theory (DFT) in the generalized gradient approximation (GGA) with the Perdew-Burke-Ernzerhof parametrization for the exchange-correlation functional \cite{perdew1996generalized}. The Kohn-Sham equations were solved by using the plane-wave pseudopotential method as implemented in the Quantum ESPRESSO software  \cite{giannozzi2009quantum}. The plane-wave energy cutoff was $47.5$ Ha for the wave function and $237.5$ Ha for the charge density. The $k$-point sampling of the Brillouin zone was done with an $5\times5\times5$ grid following the Monkhorst-Pack scheme \cite{monkhorst1976special} and with a Marzari-Vanderbilt smearing width of $5\cdot 10^{-4}$ Ha \cite{marzari1999thermal}. The crystal structure was optimized at each volume of the unit cell. During the ionic relaxation, all positions were relaxed until Hellmann-Feynman forces were less $0.05$ Ha/Bohr and the total energy converged below to $5\cdot 10^{-6}$ Ha with respect to the Brillouin zone integration.
The equation of state (EoS) of the KNaCuSi$_4$O$_{10}$ compound was obtained by fitting the total energy as a function of volume to the third-order Birch-Murnaghan equation \cite{birch1947finite}:
\begin{eqnarray}
E(V) = E_0 + \frac{9V_0B_0}{16}
\left\{ \left[\left(\frac{V_0}{V}\right)^\frac{2}{3}-1\right]^3B_0^\prime + \right. \nonumber \\
\left. + \left[\left(\frac{V_0}{V}\right)^\frac{2}{3}-1\right]^2 \left[6-4\left(\frac{V_0}{V}\right)^\frac{2}{3}\right]\right\}
\label{energy}
\end{eqnarray}
where $V_0$, $B_0$ and $B'_0$ are equilibrium volume, bulk modulus at ambient pressure and its pressure derivative, respectively. The values found for these parameters were the followings: $V_0=3271$Bohr$^3$, $B_0=54.1$GPa and $B'_0=3.3$. Figure \ref{fig:dft} shows the structural optimization curve, i.e. results for the total energy versus volume of unit cell obtained for DFT calculations and fitted by the third-order Birch-Murnaghan equation of state (Eq. (\ref{energy})).
 
For a Heisenberg spin-$1/2$ dimer there are two eigenvalues of energy, one being $E_T$ the triplet state and the other $E_S$ being the singlet state of the dimeric unit. Hence, its magnetic coupling constant ($J$) is the difference between these two states $J = E_S - E_T$. Thus by calculating the energies in Eq. (\ref{energy}) for each configuration it is possible to obtain the magnetic coupling for different values of pressure using first principles calculations.

\subsection{Results}

\begin{figure}[htp]
\centering
\subfigure[]{\includegraphics[scale=0.3]{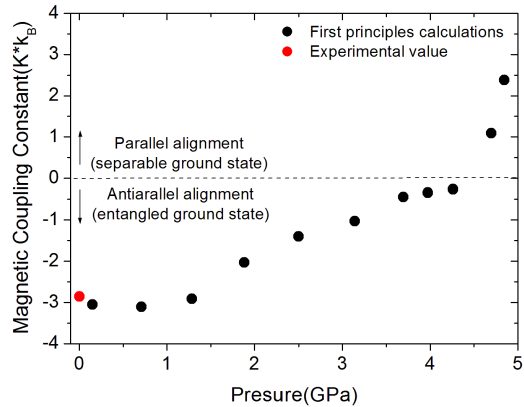}}\qquad
\subfigure[]{\includegraphics[scale=0.3]{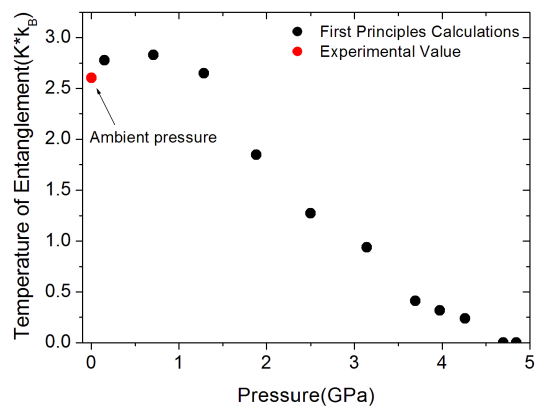}}
\caption{(colour online)(a) Magnetic coupling constant and (b) temperature of entanglement ($T_e$) obtained as a function of the external pressure. Red circle is the experimental value obtained at ambient pressure taken from reference \onlinecite{brandao2009magnetic}.
The external pressure induces a structural contraction on the prototype material leading to a change of the magnetic alignment of the system. This change yields a decrease on the degree of entanglement by reducing $T_e$.}
\label{fig:fig3}
\end{figure}

The external pressure applied on the prototype material KNaCuSi$_{4}$O$_{10}$ induces a structural contraction that reduces its lattice parameters. As a consequence, its magnetic coupling constant ($J$) increases and becomes positive, i.e., the system changes from an antiparallel alignment ($J<0$ - entangled ground state) to a parallel alignment ($J>0$ - separable ground state), as can be seen at Figure \ref{fig:fig3} (a). Hence, we obtain the temperature of entanglement ($T_e$), Eq. (\ref{temp_ent}), using the magnetic coupling constant. Figure \ref{fig:fig3} (b) shows the dependence of $T_e$ in the prototype compound under pressure. As can be seen, the change on magnetic alignment leads to decrease $T_e$ down to the disappearance of the entanglement (see Figure \ref{fig:fig3} (b)), when the system achieves a parallel alignment. This means that the degree of entanglement in a magnetic system can be controlled by significant macroscopic effects when applying external pressure.

On the other hand, in order to evaluate the influence of  external pressure on the quantum correlations, we calculate the magnetic susceptibilities (Eq.\ref{eq:4}) from each magnetic coupling constant presented in Figure \ref{fig:fig3} (a). Using Eq.(\ref{eq:6})-(\ref{eq:9}), it is possible to evaluate the quantum correlations as a function of these susceptibilities. We obtain the quantum correlation curves measured by this thermodynamic property for each magnetic configuration of the system. Thus, we establish a relationship between the quantum properties with significant macroscopic effects.

\begin{figure}[htp]
\centering
\subfigure[]{\includegraphics[scale=0.25]{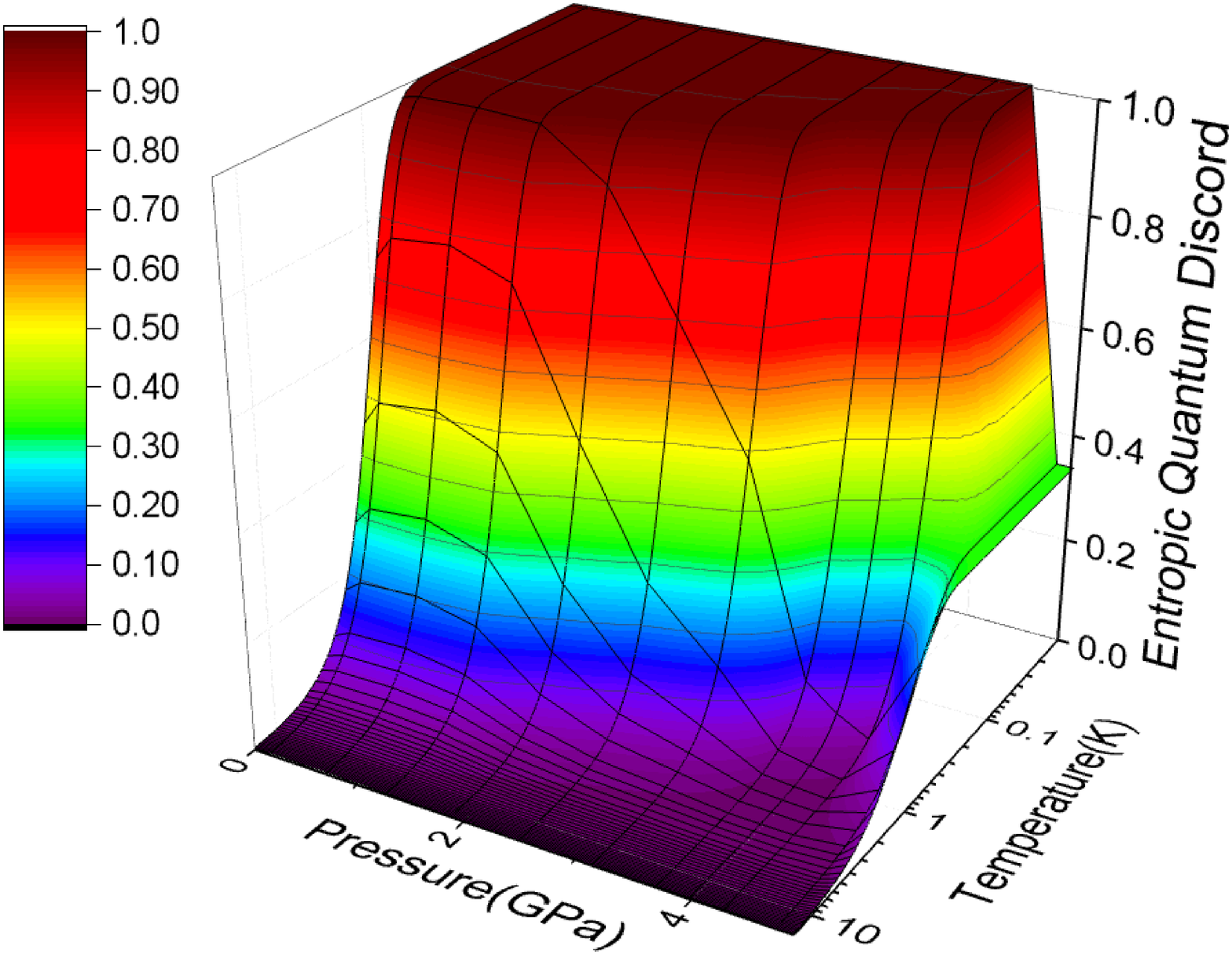}}\qquad
\subfigure[]{\includegraphics[scale=0.25]{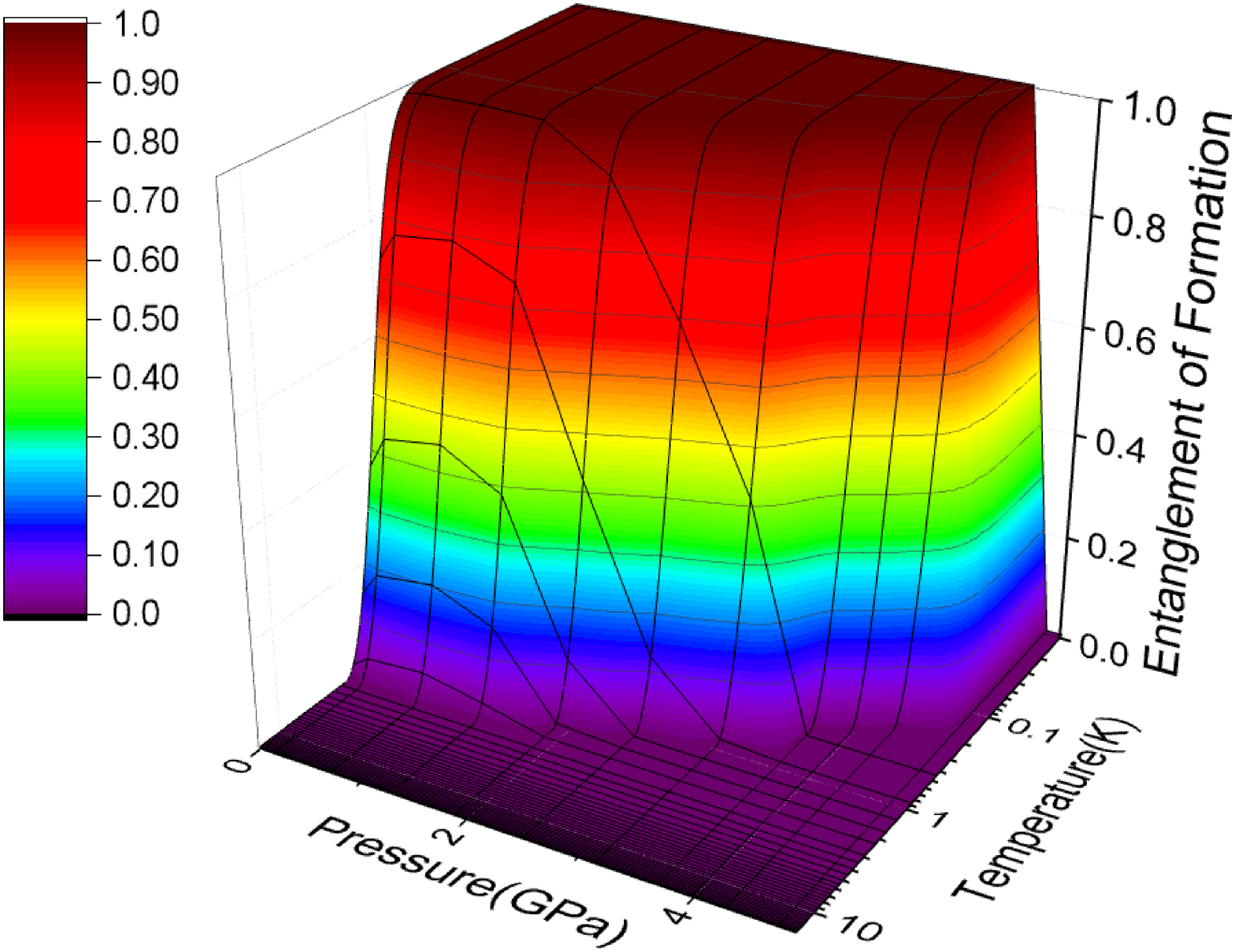}}\qquad
\caption{(colour online) (a) Entropic and (b) entanglement of formation as a function of the temperature and  pressure. It is worth to note that, it is possible to manage the degree of quantum correlation in a magnetic system  by the control of external pressure and temperature.}
\label{fig:fig1}
\end{figure}

In sequence, Figure \ref{fig:fig1} (a) and (b) shows the entropic quantum discord (Eq.\ref{eq:6}) and entanglement of formation (Eq.\ref{eq:9}) curves as a function of the temperature and pressure, respectively.  Note that the degree of quantum correlations in the system decreases by increasing the external pressure as a consequence of the changes of the magnetic coupling constant as shown in Figure \ref{fig:fig3} (a). Also remarkable is the management of the ground state in the system by reducing the temperature and controlling the external pressure; it achieves this state when the entropic discord (Figure \ref{fig:fig1} (a)) and entanglement (Figure \ref{fig:fig1} (b)) reach the maximum value of unity. As pointed out before, the system changes from the entangled ground state to a separable ones due to the changes of the magnetic configuration of the system induced by the external pressure. It is reflected in the quantum correlations when entanglement goes to zero (see Figure \ref{fig:fig1} (b)), i.e., system reaches the temperature of entanglement  $T=T_e$, Eq. (\ref{temp_ent}). Hence, this fact yields a minimization on the quantum discord in the system (see Figure \ref{fig:fig1} (a)). However, it remains significantly different of zero even in separable states when the entanglement is absent as can be seen at Figure \ref{fig:fig1} (a). In this way, we can control the quantum correlations of this molecular magnet by means of macroscopic properties such as temperature and pressure.

Therefore, the  structural contractions on the system, achieved by increasing the external pressure, leads to a change in the magnetic configuration of the system, as can be seen at Figure \ref{fig:fig3}. As a consequence, the quantum correlations (Figure \ref{fig:fig1}) are drastically affected, since these quantum properties are directly related to the magnetic behaviour of the system (Eq.(\ref{eq:6})-(\ref{eq:9})). Thus, it is possible to manage the degree of quantum correlation in a magnetic system  by the control of external pressure and temperature.
Furthermore, this external pressure can be achieved experimentally reducing the lattice parameter of the system by chemical substitution or hydrostatically, for example. This fact opens a large avenue for research in experimental detection and manipulation of quantum correlations. It allows a better understanding of the quantum properties of molecular magnets by the management of significant macroscopic properties, leading to promising applications in quantum information science such as the development of novel candidate platforms  for quantum information processing by means of materials engineering.

\section{Conclusions}

In summary, we performed first-principles calculations to investigate the dependence under hydrostatic pressure, of the quantum correlations of the prototype material  KNaCuSi$_{4}$O$_{10}$, which is a Heisenberg dimer in a $d^9$ electronic configuration and, consequently, an ideal realization of a two qubit system (spin $1/2$ dimer). We show that an external pressure induces a structural contraction on the prototype material, leading to a change of its magnetic alignment. This changes yields a minimization on the degree of the quantum correlations in the system; showing that quantum correlations are related to significant macroscopic effects. Also remarkable is the possibility to handle the ground state by controlling the temperature and pressure. Our results allow a better management of the quantum properties in magnetic systems and opens the doors for  experimental and theoretical research of quantum correlations in these systems via first-principles, leading to a better understanding of the quantum properties of these systems with promising applications in emerging quantum technologies.
\begin{acknowledgments}
The authors would like to thank the Brazilian funding agencies CNPq, CAPES and FAPERJ. This work was developed within the support of CENAPAD-UNICAMP.
\end{acknowledgments}

\end{document}